# A catalog of intermediate duration Type I X-ray bursts observed with the INTEGRAL satellite

K. Alizai[1]⋆, J. Chenevez[1]†, S. Brandt[1], N. Lund[1],

[1] *National Space Institute, Technical University of Denmark,*

20 March 2020

**ABSTRACT**
We present a catalog of long duration bursts observed with the Joint European X-ray Monitor (JEM-X) and IBIS/ISGRI instruments onboard the INTEGRAL satellite. The fourteen bursts have e-folding times ranging from 55 s to ≈ 17 min, and are therefore classified as intermediate-duration bursts, caused by the ignition of an unusually thick helium layer. Though seven events have already been reported in literature, we have systematically re-analyzed the whole sample. We find three new photospheric radius expansion (PRE) bursts, which are not reported in the literature, allowing us to provide a new estimate of the distances to these sources. We apply the enhanced persistent emission method (also known as the $f_a$ method) on sources with detectable persistent emission prior to a burst, in order to follow the evolution of the accretion rate during the burst. Although we do not get significantly better fits, the evolution of the $f_a$ factor shows an indicative behavior, which we discuss.

**Key words:** X-ray binaries: X-ray bursters – stars: – stars: neutron – X–rays: bursts

## 1 INTRODUCTION

More than 110 X-ray bursters have been identified in the Milky Way[1] since the first thermonuclear X-ray bursts were observed in 1975 (Grindlay et al., 1976; Swank et al., 1977). These are a subclass of X-ray binaries composed of a weakly magnetized neutron star, that accretes matter (hydrogen and/or helium) through the Roche lobe overflow from an orbiting lower mass companion star. In the most common scenario, the accumulated stellar matter on the surface of the neutron star forms an upper layer of hydrogen, which, due to the high temperature and density, steadily burns through thermonuclear fusion into a sublayer of helium. For some accretion rates, the burning layer reaches a critical thickness above which the nuclear reactions become unstable at temperatures exceeding 20 million K.

Depending on their duration, X-ray bursts are divided into three groups.

• The most common X-ray bursts (≈ 99%) have recurrence times of a few hours. They are caused by ignition of either a pure helium (≈ 1 m thick) layer or a mixed helium and hydrogen layer. Mostly depending on the thickness of the burning layer, the duration of the bursts can be from 10$s$ to a few 100s seconds (Maraschi Cavaliere, 1977; Lewin, van Paradijs & Taam, 1993; Strohmayer & Bildsten, 2006 & Galloway & Keek, 2017); > 8000 bursts have been recorded in the past 50 years (Galloway, in 't Zand & Chenevez, 2017). For the pure helium bursts, the luminosity can often reach the local Eddington limit, causing a photospheric radius expansion (PRE) burst. The accretion luminosity of the sources, prior to a burst, is around a few percent of the Eddington luminosity.

• X-ray bursts with durations of few minutes up to a few 10s of minutes are called intermediate-duration bursts. They have recurrence time of weeks to months and are due to the ignition of a thick (10 − 100 m) layer of helium. Most commonly, the accretion luminosity prior to an intermediate-duration X-ray burst is ≈ 1% of the Eddington limit, GX 17+2 being the exception (Kuulkers et al., 1997, Kuulkers et al., 2002). To date, 70 of these bursts have been observed.

• The rarest X-ray bursts (27 to date) have durations of several hours up to a day and are called superbursts. Simulations indicate that superbursts have recurrence times of a year (Keek et al., 2008) and have an energy output $10^3$ times greater than that of regular bursts. Superbursts occur for a wide range of accretion rates and are usually sub-Eddington (although precursors to superbursts do reach the Eddington limit). They are thought to be due to an ignition of a deep carbon layer.

---
⋆ E-mail: kha@space.dtu.dk   † E-mail: jerome@space.dtu.dk
[1] https://personal.sron.nl/ jeanz/bursterlist.html, (last update; April 26$^{th}$, 2018)

To date, we have only observed a single superburst with the INTEGRAL/JEM-X instrument from SAX J1747.0-2853 on February 11, 2011 (to be published in a separate paper).

The BeppoSAX Wide Field Camera instrument (Jager et al., 1997) and the Joint European X-ray Monitor (JEM-X) on the INTEGRAL satellite are, because of their large field of view (FOV) (0.3% of the sky for INTEGRAL/JEM-X and 8% for BeppoSAX/WFC), the utmost instruments for detecting long duration bursts. These are both coded-mask telescopes, that image by blocking radiation in a known pattern casting a "shadowgram" on the detector plane, which is then reconstructed to an image (in 't Zand et al., 2002).

Because of their long duration, superbursts can also be detected by instruments that do not have a particularly large FOV, but that scan the sky on regular intervals. Such instruments are the ISS/MAXI and the RXTE/ASM that scan > 80% of the sky every 1.5 hours.

There are other instruments in orbit that have a large FOV, but their band passes do not cover the relevant energy range for X-ray bursts and at most these devices can detect the hard X-rays from the bursts, which are evident in the beginning of the bursts, but fade out quickly. The peak effective temperature of an X-ray burst, derived from the spectroscopy, is around $2 - 3$ keV. This indicates that the energy spectrum peaks at $6 - 9$ keV. Amongst the wide-field cameras most preferable for the hardest photons are IBIS/ISGRI detector onboard INTEGRAL (18 keV - 10 MeV) (Chernyakova et al., 2015), the imaging Burst Alert telescope (BAT) onboard the Swift observatory (15 − 150 keV) (Markwardt et al., 2007) and the Fermi Gamma-ray Burst Monitor (GBM: 8 keV - 1 MeV) (Band & Ferrara, 2009).

Thermonuclear X-ray bursts are, in general, interesting for observers, because they can improve our understanding of the fuel composition, accretion rate, as well as the neutron star spin, mass and radius (Heger et al., 2007; Strohmayer & Bildsten, 2006). The rare intermediate-duration bursts are particularly interesting, because they can be used to probe the thermal properties of the neutron star interior, as they occur deep in the neutron star envelope and their ignition is therefore sensitive to the thermal profile of the crust and the core of the neutron star (Cumming et al., 2006). Different numerical models have been used to understand the burning physics, ignition conditions and energy release of intermediate-duration bursts, but the model predictions have not been compared to observations in a sufficient degree, partly because of the lack of a statistically significant sample (Falanga et al., 2008; in 't Zand et al., 2005; Kuulkers et al., 2010).

In this paper we present intermediate-duration X-ray bursts detected with instruments onboard the INTEGRAL satellite. In section 2 we give an overview of the INTEGRAL/JEM-X instrument and the data analysis software. In section 3 we present the data sample used for this study and give an overview of the data from literature, which has been reused for this study. In section 4 we present our data analysis methodology. Section 5 is dedicated to a presentation of our results. In section 6 a discussion of our results and the physics that can be done with a catalog like ours. In section 7 we provide a summary and a conclusion.

## 2 INSTRUMENTATION AND DATA ANALYSIS SOFTWARE

The INTEernational Gamma-Ray Astrophysics Laboratory (INTEGRAL) was launched into orbit in 2002 by the European Space Agency (ESA). INTEGRAL is a dedicated 20 keV - 10 MeV gamma-ray observatory with secondary instruments that have the capability of source monitoring at X-rays (JEM-X, $3 - 35$ keV) and in the optical range of the electromagnetic spectrum (OMC, $500 - 600$ nm). The primary instruments of the INTEGRAL payload are the SPECTROMETER (SPI; 15 keV - 10 MeV and the IMAGER (IBIS: 20 keV - 10 MeV), the latter consists of two detector planes: ISGRI and PICsIT.

The instruments of interest for our study is JEM-X and the ISGRI detector. JEM-X consists of two twin coded-mask telescopes, which are co-aligned with the other instruments on INTEGRAL. JEM-X has a useful FOV of $5.5^o$ in diameter, $3'$ angular resolution and an energy resolution of $\Delta E/E = 47\%(E/1keV)^{-1/2}$. The energy range of JEM-X is $3 - 35$ keV (Lund et al., 2003; Kuulkers, 2011). ISGRI is the top layer of the IBIS instrument. It has a collecting area of 2600 cm$^2$, has an energy resolution of 8%@100 keV, angular resolution of $12'$ FWHM, FOV of $8.3^o \times 8.0^o$ (fully coded), and an energy of $15 - 1000$ keV (with maximum effective area between 20 and 100 keV) (Savchenko et al., 2018).

A single pointing of INTEGRAL is called a science window (ScW). For this study, we have used the standard off-line Science Analysis (OSA) software version 11. OSA11 for JEM-X has 11 scientific levels, through which JEM-X data is corrected for instrument effects and deadtime. Furthermore, Good Time Intervals (GTI) are created to be used for image reconstruction and spectra extraction (Westergaard et al., 2003).

## 3 OBSERVATIONS

The observations included in this study were carried out over the span of fourteen years (2004 - 2018). So far, 15 intermediate-duration bursts have been serendipitously detected from 8 sources by INTEGRAL, since its launch. 11 bursts are only detected by JEM-X, three bursts are detected by both JEM-X and IBIS/ISGRI, and one burst is detected by IBIS/ISGRI alone. nine of the bursts have been previously published in peer-reviewed papers or in Astronomer's Telegrams (ATel). Table 1 shows a list of the bursts included in this study. We present light curves and time-resolved spectral analyses (TRSA) for 14 of the 15 intermediate-duration bursts detected by INTEGRAL, where we have re-done the TRSA for 5 bursts with different time intervals and different models relative to those, published in the literature. The missing burst is from the source SLX 1735-269 detected in $15^{th}$ of September, 2003, during which JEM-X was running in restricted mode (Molkov et al., 2005), a data format that is not supported by OSA 11. All uncertainties reported in this paper are at $1\sigma$.

**Table 1.** List of all long X-ray bursts investigated in this study.

| Source | detection date | MJD | Revolution/ScW | Off-axis($^o$) | Instrument(s) | Reference |
|---|---|---|---|---|---|---|
| SLX 1737-282 | 2004-03-09 | 53073.72296 | 0171/75 | 5.05 | JEM-X 1 | Falanga et al., 2008 |
| SLX 1735-269 | 2004-04-13 | 53108.20530 | 0183/11 | N/A | IBIS/ISGRI | Sguera et al., 2007 |
| GX 3+1 | 2004-08-31 | 53248.78751 | 0230/09 | 2.46 | JEM-X 1 | Chenevez et al., 2006 |
| SLX 1737-282 | 2005-04-11 | 53471.35022 | 0304/61-62 | 1.72/1.74 | JEM-X 1 | Falanga et al., 2008 |
| AX J1754.2-2754 | 2005-04-16 | 53476.92472 | 0306/50 | 1.85 | JEM-X 1 | Chelovekov & Grebenev, 2007 |
| IGR J17254-3257 | 2006-10-01 | 54009.30180 | 0484/42-43 | 3.66/2.76 | JEM-X 1 | Chenevez et al., 2007 |
| SLX 1737-282 | 2007-04-02 | 54192.24886 | 0545/48 | 3.73 | JEM-X 1 | Falanga et al., 2008 |
| GX 17+2 | 2012-03-25 | 56011.77901 | 1153/116 | 2.03 | JEM-X 1 & 2 | This work |
| GX 17+2 | 2012-08-21 | 56160.00084 | 1203/49 | 2.93 | JEM-X 1 & 2 | This work |
| SLX 1744-299 | 2013-04-06 | 56388.46546 | 1279/69 | 2.66 | JEM-X 1 & 2 | This work |
| SLX 1744-299 | 2015-02-27 | 57080.41900 | 1512/58 | 2.42 | JEM-X 1 & 2 | This work |
| SLX 1744-299 | 2016-03-07 | 57454.43753 | 1653/16 | 3.41 | JEM-X 1 & 2 | This work |
| AX J1754.2-2754 | 2017-03-12 | 57824.16536 | 1792/17 | 1.17 | JEM-X 1 & 2 | This work |
| SAX J1712.6-3739 | 2018-02-13 | 58169.96485 | 1922/13 | 1.29 | JEM-X 1 & 2 | This work |

## 4 ANALYSIS OF THE DATA

The JEM-X light curves can be extracted for several different energy bands in the detector energy range of $3-35$ keV. Since most of photons from a bursting source have energies below 25 keV, we extract light curves between $3-25$ keV. We fit the light curves with an exponential function (and a constant component for the noise), and quote the e-folding time as $\tau$.

Due to the relatively small effective area of JEM-X($60-100$ cm$^2$, E-depending), we use only eight energy bins to extract the spectra. The choice of time intervals for the TRSA is based on the consideration of having a minimum of 1200 counts per spectrum in our TRSA. This minimum is chosen to ensure enough counts in each of the energy bins, as a burst progresses and its spectrum becomes softer. Therefore, the length of the intervals is increased through the evolution of the burst.

The spectral analysis is performed with the XSPEC software (Arnaud, 1996). Because of the low number of energy bins, we use simple fitting models in an attempt to have the highest number of degrees of freedom (dof) as possible (dof = 4 for all the spectra included in this study). We set the solar abundance and the photoionization cross-section to WILM and VERN (in XSPEC syntax), respectively. We then use the model TBABS as our absorption model. We quote the absorption column from literature and freeze the value. The persistent emission of the sources is only detected for 7 out of the 14 bursts we have analyzed. We model the persistent emission with a simple powerlaw (PO in XSPEC syntax) for 5 of the 7 cases. For the remaining 2 cases, we model their persistent emission with the XSPEC model CUTOFFPl, which is a powerlaw with high energy exponential rolloff. The models are chosen so they give the best fit for our data. We then include the model for the persistent emission in the model for the burst itself, but we freeze all model parameters from the persistent emission model. We assume that the bursts themselves emit as a blackbody reasonably well and use the XSPEC model BBODYRAD to fit the burst emission component. For bursts without any detectable persistent emission, we fit the spectra with the blackbody model BBODYRAD only.

The distances used in this study are either quoted from the literature or estimated from PRE-bursts. We have looked through the Data Release 2 of the Global Astrometric Interferometer for Astrophysics (Gaia) for distances measured to the sources of interest for our study, since most of our sources are in crowded regions of the sky, we do not get any reliable parallax for them and we can therefore not quote any Gaia distances. We use the source-distances as fixed values without considering their uncertainties.

## 5 RESULTS

### 5.1 Burst light curves

We present the count-rate light curves of the 14 X-ray bursts detected by INTEGRAL in fig. A1, appendix A. Fig. A1[(a)-(m)] show the JEM-X light curves (3 - 25 keV), where the bottom panel of fig. A1[(h)-(j)] show the IBIS/ISGRI light curves (20 - 60 keV) of the bursts. Fig. A1[(n)] show IBIS/ISGRI light curve (20 - 60 keV) of the 2004 burst from SLX 1735-269 detected only by IBIS/ISGRI. The e-folding times, obtained from fitting the count-rate light curves with an exponential function, for 13 bursts (the only IBIS/ISGRI burst not included) range from $55 \pm 10$ s up to $1011 \pm 140$ s. The e-folding times are listed in table 2. We observe the rise phase of 12 bursts in our sample. The Two-phase X-ray burst from GX 3+1 is the only burst in our sample with a fast rise time ($< 2$ s), the remaining bursts all exhibit long rise times ($> 10$ s). We choose to plot each light curve with a time-bin that reduces noise as much as possible.

### 5.2 Bursts from AX J1754.2-2754

The first observation of the source AX J1754.2-2754 was made by the ASCA observatory on October 2-3, 1999 (Sakano et al., 2002), but it was not until 2005, that the nature of the compact object could be determined, when the first burst was detected (Chelovekov & Grebenev, 2007). The interstellar absorption of the source is found to be $N_H = 2.0 \times 10^{22}$ cm$^{-2}$ (Sakano et al., 2002).

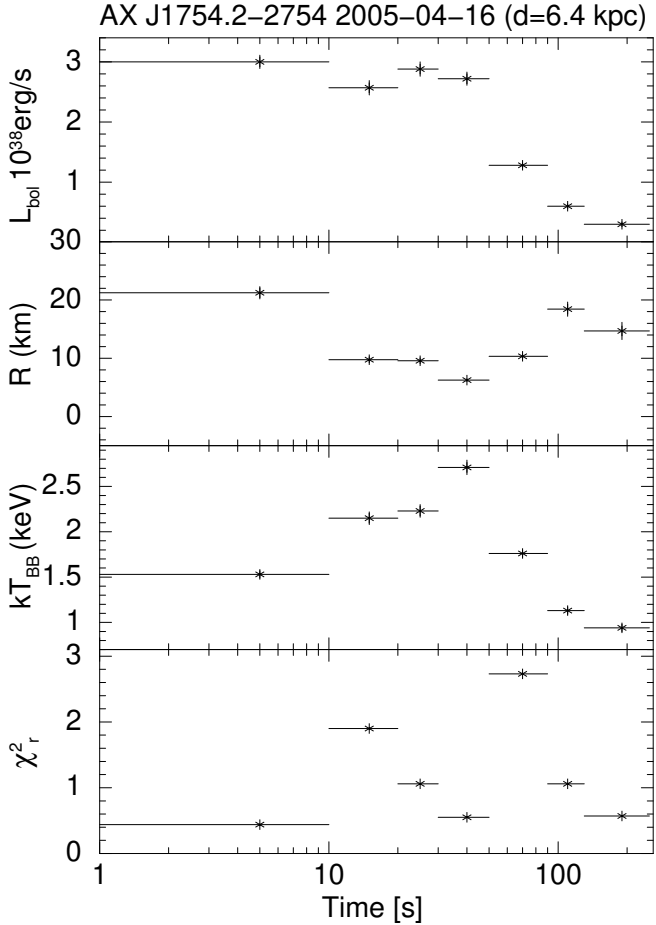

**Figure 1.** AX J1754.2-2754 / 2005-04-16: Time-resolved spectroscopy.

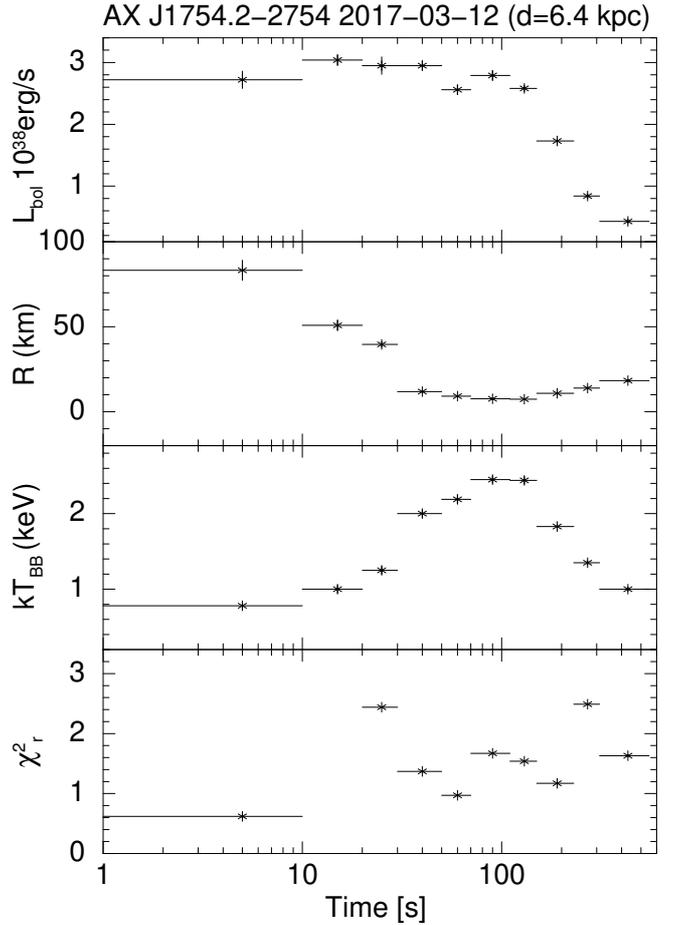

**Figure 2.** AX J1754.2-2754 / 2017-03-12: Time-resolved spectroscopy.

*5.2.1 April 16, 2005*

This burst was already published by (Chelovekov & Grebenev, 2007). Based on this burst, the source was classified as an X-ray burster. Furthermore, two distances were estimated to the source, since it clearly was a PRE burst. The two distances estimated to the source were $d = 6.6 \pm 0.3$ kpc (for hydrogen atmosphere) and $d = 9.2 \pm 0.4$ kpc (for helium atmosphere). For the latter the empirically obtained Eddington luminosity $L_{Eddington} = 3.8 \times 10^{38}$ erg/s was used (Kuulkers et al., 2003). We have reestimated the distance to the source using the theoretical value of $3.0 \times 10^{38}$ erg/s for the He Eddington luminosity (assuming a $1.4 M_\odot$ neutron star with $R_{NS} = 10$ km), and derive a distance of $d = 7.01 \pm 0.2$ kpc, at a peak flux of $5.2 \times 10^{-8}$ erg cm$^{-2}$ s$^{-1}$ (see fig. 1).

*5.2.2 March 12, 2017*

The 2017 burst from AX J1754.2-2754 is the longest burst detected from the source. Like the burst from 2005, the 2017 burst is also a PRE burst. We estimate the distance to the source based on the peak flux of this burst also and get 7.2 kpc. With the peak flux of $4.9 \times 10^{-8}$ erg cm$^{-2}$ s$^{-1}$, the distance is derived (using $3.0 \times 10^{38}$ erg/s as the Eddington luminosity). This is the most recent burst detected from AX J1754.2-2754 (see fig. 2).

### 5.3 Bursts from GX 17+2

The bright persistent source GX 17+2 is only one of two Z-sources that have showed X-ray bursts, the other being Cygnus X-2. GX 17+2 accretes matter close to the Eddington limit, and has shown short-, intermediate X-ray bursts and superbursts. A distance of 12 kpc has been derived for the source from 5 PRE bursts (Kuulkers et al., 2002, in 't Zand et al., 2004). The JEM-X instrument has detected two intermediate duration X-ray bursts from GX 17+2. Both of them occurred in 2012 almost 5 months apart.

*5.3.1 March 25, 2012*

The first intermediate burst from GX 17+2 was detected on the $25^{th}$ of March, 2012. The burst onset occurs 120 s after the start of the ScW. The tail of the burst seems to continue even after it is not in the FOV of the instrument. The peak flux of the burst is $5.0 \times 10^{-8}$ erg cm$^{-2}$ s$^{-1}$, which corresponds to a peak luminosity of $8.62 \times 10^{38}$ erg s$^{-1}$, which is well above the Eddington luminosity. We can instead use the distance calculated by (Kuulkers et al., 2002), where they considered the gravitational redshift effects and estimated a distance of 8 kpc. The results presented in fig. 3 are calculated assuming a distance of 8 kpc. The blackbody temperature for most of the burst is above 2 keV. Notice

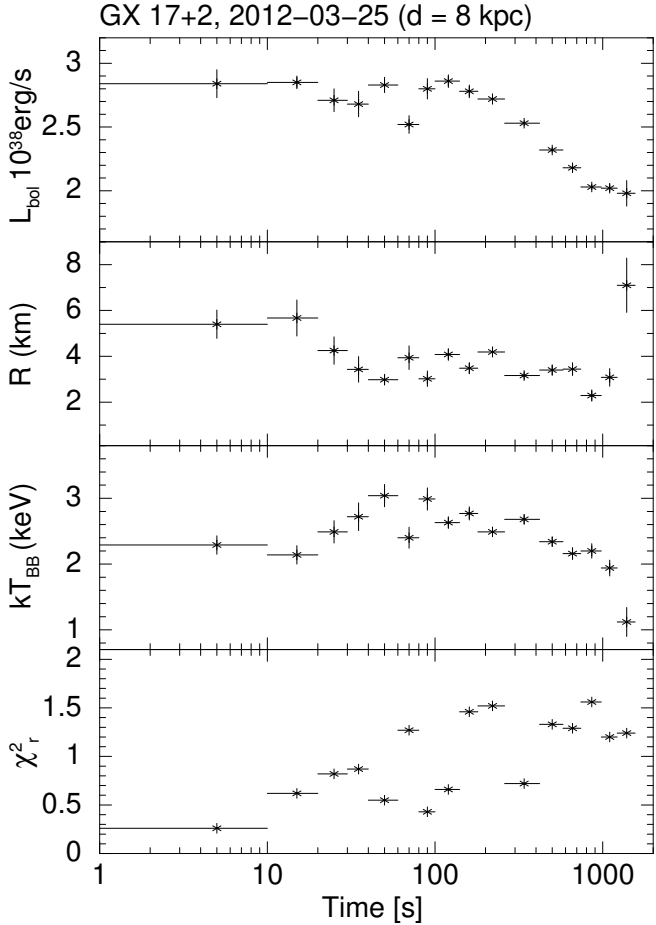

**Figure 3.** GX 17+2 / 2012-03-25: Time-resolved spectroscopy.

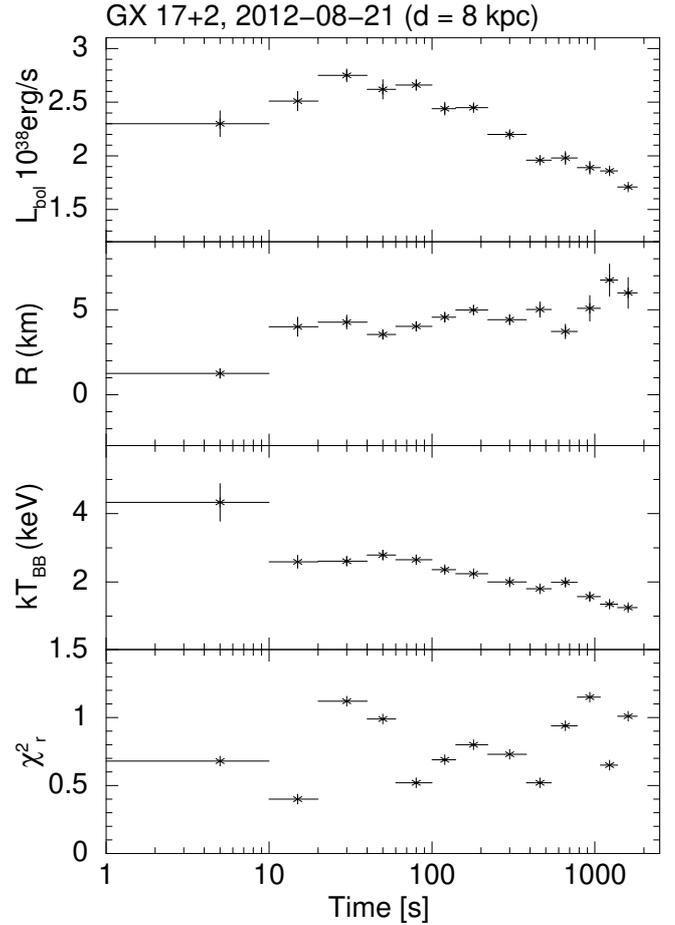

**Figure 4.** GX 17+2 / 2012-08-21: Time-resolved spectroscopy.

the significant decrease in the temperature at the last data point.

*5.3.2 August 21, 2012*

The second JEM-X intermediate burst from GX 17+2 was detected on the $21^{st}$ of August, 2012. The peak of the burst is not detected, since the source is not in the FOV at the time. We see the tail of the burst from the start of the ScW and continues to decrease in flux throughout the entire of ScW. As for the March-burst, the blackbody temperature is above 2 keV for the large part of the burst and we first notice a significant decrease in the latter part of the burst (see fig. 4).

### 5.4 Burst from GX 3+1

GX 3+1 is a persistent atoll source, which is always in the soft (banana) state (Seifina & Titarchuk, 2012). The source is a well-known burster that has shown regular bursts and a superburst (Kuulkers, 2002). INTEGRAL/JEM-X detected an unusual burst from GX 3+1 on $31^{st}$ of August, 2004. A time-resolved spectral analysis of the burst was presented by (Chenevez et al., 2006), which clearly showed spectral softening. The unusual aspects of the burst from 2004 are the initial spike and the prolonged tail, which continues for more than 2000 s. (Chenevez et al., 2006) used a two component spectral analysis for both the persistent emission and burst emission. The model used contained a blackbody (BB) component for the thermal emission and the power-law (PO) component for the Comptonized photons.

We have reanalyzed the burst using also a two-component spectral model, but we use two different models for the persistent emission and the burst emission, respectively. We use the DISKBB model (with the absorption model TBABS) from XSPEC for the persistent emission, where we assume an accretion disk consisting of multiple blackbody components. For the burst emission we use an absorbed blackbody (BBODYRAD). We use $N_H = 1.6 \times 10^{22}$ atoms $cm^{-2}$ derived by (Oosterbroek et al., 2001). For the model parameters and other burst parameters of the whole burst-sample, see Table 2. Although we do not use the same models as in the original publication, we do get the same parameter values (see fig. 5). But if we use a PO to model the persistent emission, we get poorer fits than we do with DISKBB model.

### 5.5 Burst from IGR J17254-3257

IGR J17254-3257 was discovered by INTEGRAL in 2003 in the Galactic Centre hard X-ray survey (Walter et al. (2004)). To date, only two X-ray bursts have been detected from this source. The first X-ray burst was an ordinary one, detected on February $17^{th}$, 2004. The second burst was detected on October $1^{st}$, 2006 and was identified as an intermediate burst (Chenevez et al., 2007). We present the first ever time-resolved spectral analysis of the second burst from this source in figure 6. There is a data gap of 2 min during the burst, caused by the slew of the instrument. We use 14.5 kpc as the upper limit of the distance, derived by using $3.8 \times 10^{38}$ erg s$^{-1}$. We use $N_H = 1.79 \times 10^{22}$ cm$^{-2}$ which was obtained as one of the best fit parameters by Chenevez et al., 2007 using the joined spectrum of pn/MOS1/JEM-X/ISGRI. The source is too weak to be detected prior to the burst, which prevents us from using the two component method, and instead we fit the burst spectra with an absorbed Black body.

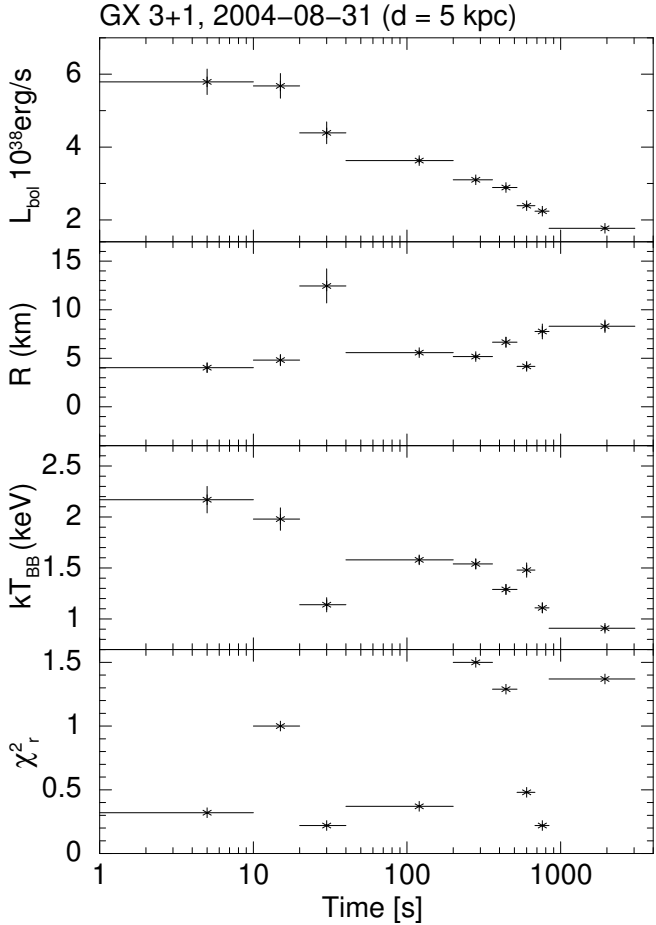

**Figure 5.** GX 3+1: Time-resolved spectroscopy.

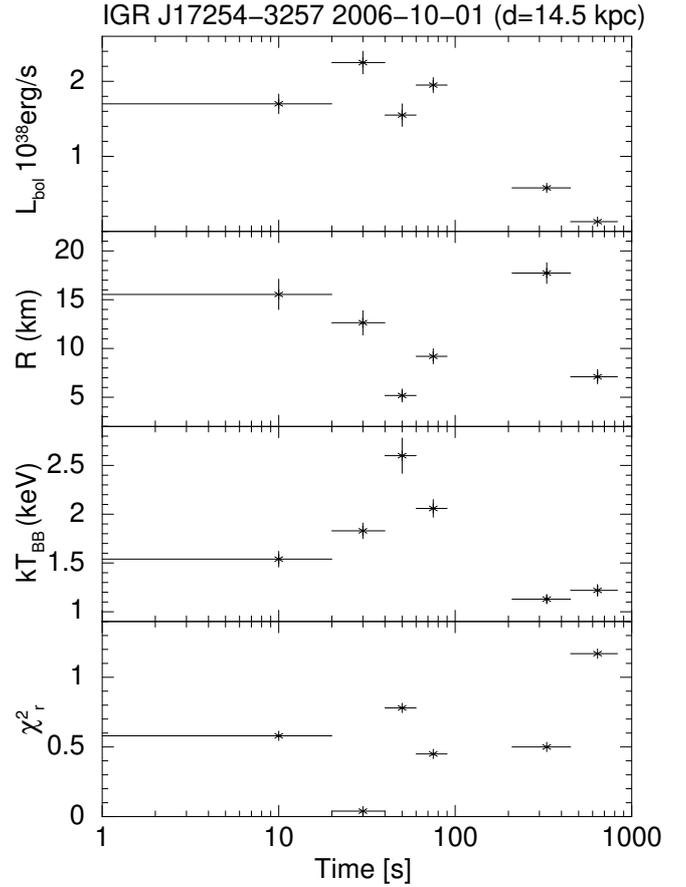

**Figure 6.** IGR J17254-3257: Time-resolved spectroscopy.

### 5.6 Burst from SAX J1712.6-3739

SAX J1712.6-3739 source was discovered by BeppoSAX in 1999 (in 't Zand et al., 1999), and it has been categorized as a persistent source since 2001. Immediately after its discovery, the source showed two regular Type I X-ray burst. Based on these first X-ray bursts, the distance to the source was estimated to 7 kpc (in 't Zand et al., 1999). To date, the source have shown a several intermediate bursts and a single superburst-like flare, detected with the Burst Alert Telescope (BAT) onboard the Swift satellite. The only intermediate X-ray burst, detected by JEM-X, occurred on February 20, 2018.

We present the TRSA in fig.7. The spectra are fitted using a column density $N_H = 2 \times 10^{22}$ in 't Zand et al., 1999. Since the burst do not show a clear PRE, we assume the distance of 7 kpc. The burst reaches the peak flux of $F_{peak} = (1.1 \pm 0.2) \times 10^{-7}$, which gives us a peak luminosity of $L_{peak} = (6.45 \pm 1.17) \times 10^{38}$, well above the local Eddington luminosity. Using the peak flux, we derive an upper limit for the distance by fixing the Eddington luminosity to $L_{Edd} = 3.0 \times 10^{38}$. The result is a distance of $d = 4.77 \pm 0.8$ kpc, but we refrain from using this distance in our TRSA, since it is an upper limit.

### 5.7 Bursts from SLX 1737-282

SLX 1737-282 was discovered in 1985 with the *Spacelab - 2* (Skinner et al., 1987). The source has since been categorized as a low persistent source. Only intermediate X-ray bursts have been observed from the source (4 to date). Falanga et al., 2008 published their results on the investigation of the 3 intermediate bursts, observed with JEM-X, in 2008, where they confirmed one of the three to be a PRE burst. We have reanalyzed the data from the 3 bursts, where we have increased the number of spectra extracted for each burst. The column density in the direction of the source is fixed to $N_H = 1.9 \times 10^{22}$ for all our spectral fits. We assume a distance of 7.3 kpc, which was observationally obtained by Falanga et al., 2008.

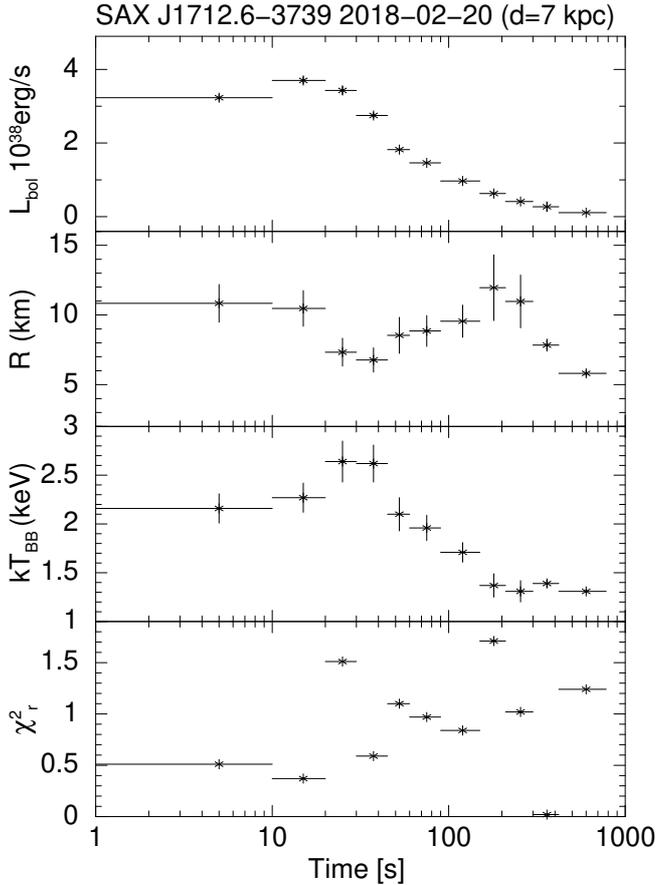

**Figure 7.** SAX J1712.6-3739: Time-resolved spectroscopy.

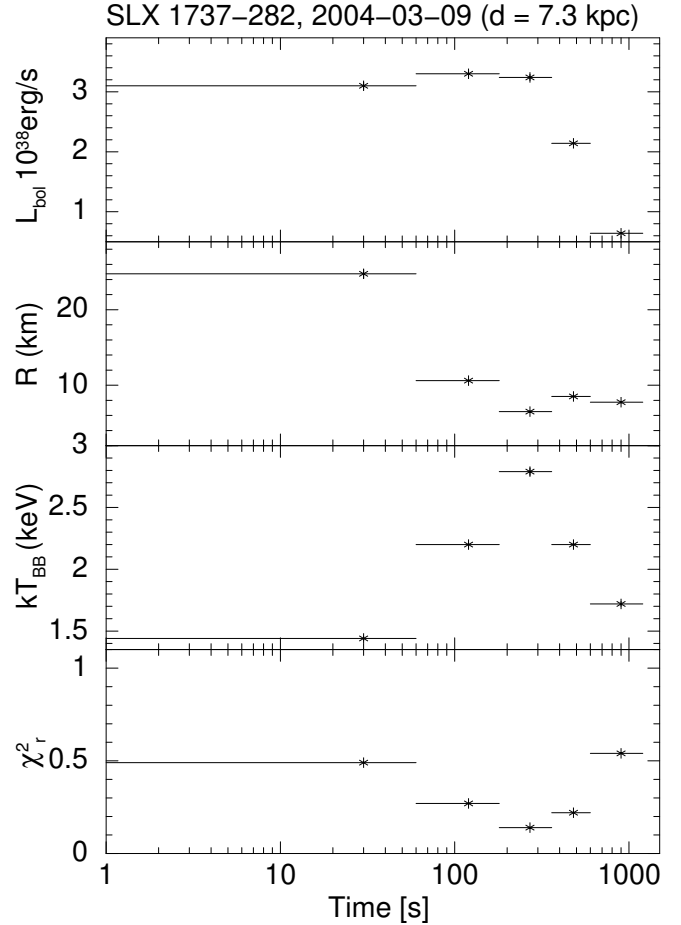

**Figure 8.** SLX 1737-282: Time-resolved spectroscopy.

*5.7.1 March 09, 2004*

This is the first burst detected by JEM-X from the source, and it is also the burst with the shortest e-folding time of the 3. Falanga et al., 2008 did not identify this burst as a PRE, but from our spectral analysis, we see an anti-correlation between the blackbody norm and the blackbody temperature $kT_{BB}$, which is indicative of a PRE. We show our TRSA in fig. 8.

*5.7.2 April 11, 2005*

The burst with the longest e-folding time of the 3 was detected in 2005. This burst was identified by Falanga et al., 2008 as a PRE, and was used to estimate the 7.3 kpc distance. Compared to their analysis, we have extracted 12 spectra of the burst instead of 8. This gives us a better detailed picture of the burst itself, where we have been able to resolve the start of the expansion phase, indicated by a decrease in the $kT_{BB}$ at the start of the burst (see fig. 9).

*5.7.3 April 02, 2007*

Just as for the first burst, this last burst from the SLX 1737-282 was not identified as a PRE. But our spectral analysis again show an anti-correlation between the blackbody norm and $kT_{BB}$ (see fig. 10), indicating a PRE-burst. To insure that this anti-correlation between the norm and the blackbody temperature is real, we fixed the temperature several intermediate values and we saw that the quality of our fits deteriorated, indicating that a variable temperature and radius is required to fit the data.

### 5.8 Bursts from SLX 1744-299

SLX 1744-299 was discovered with *Spacelab - 2* in 1987 (Pavlinsky et al., 1994). The source is close to another burster SLX 1744-300 on the sky, it was therefore only possible to measure a combined flux of the two sources with all the non-focusing telescopes. The only measurement of the fluxes of the both sources was made on the basis of a 2004 XMM-Newton observation, which resolved both sources, and found a ratio of 2.8/1.0 between SLX 1744-299 and SLX 1744-300. in 't Zand et al., 2007 categorized this source as an UCXB candidate, based on the apparent low accretion rate (i.e. slow recurrence of bursts). Nevertheless, it is interesting to note, that none of the bursts (6 intermediate bursts to date) detected from SLX 1744-299 show a PRE, even though the accreted matter is thought to be He.

We present TRSA of 3 intermediate bursts from SLX 1744-299. We assume a column density of $N_H = 4.34 \times 10^{22}$

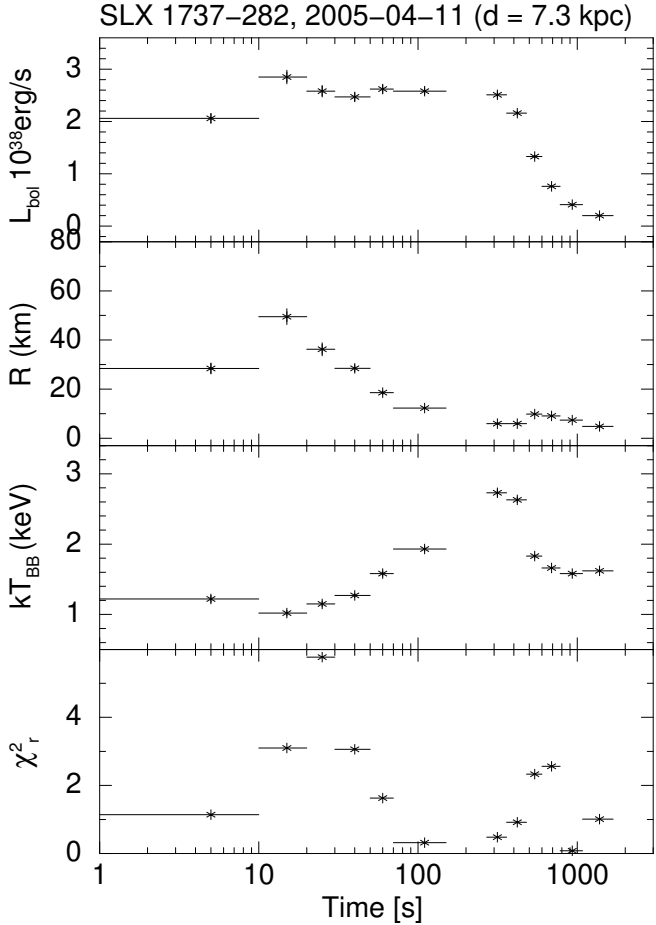

**Figure 9.** SLX 1737-282: Time-resolved spectroscopy.

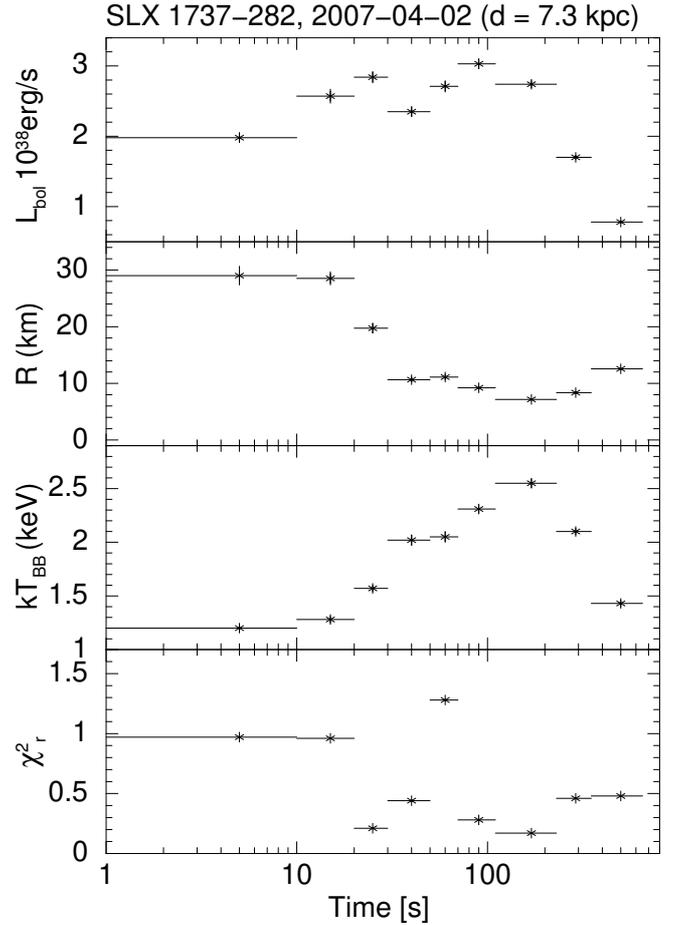

**Figure 10.** SLX 1737-282: Time-resolved spectroscopy.

cm$^{-2}$ and a distance to the source of 8.5 kpc, both values obtained from XMM-Newton observations (Pavlinsky et al., 1994).

*5.8.1 April 06, 2013*

Fig. 11 show the TRSA of the first burst detected from this source. The effective duration of the burst is approximately $\approx 200$ s, the peak flux is $F_{peak} = (4.3 \pm 0.4) \times 10^{-8}$ erg cm$^{-2}$ s$^{-1}$ and the peak blackbody temperature $kT_{BB} = 2.36 \pm 0.07$ keV (see fig. 11).

*5.8.2 February 27, 2015*

The onset of this burst occurs during the slew from one pointing to another. We therefore only see the tail of the burst, where the maximum flux registered is $F_{max} = (2.5 \pm 0.2) \times 10^{-8}$, the maximum blackbody temperature $kT_{BB} = 2.73 \pm 0.08$ keV and we have a lower limit for the effective duration of $\approx 300$ s, see fig. 12.

*5.8.3 March 07, 2016*

This burst is the shortest of all the three, observed by JEM-X. The effective duration of this burst is $\approx 100$ s, the peak blackbody temperature $kT_{BB} = 2.40 \pm 0.07$ keV and a peak flux of $F_{peak} = (3.8 \pm 0.15) \times 10^{-8}$, see fig. 13.

The three bursts from SLX 1744-299 all show a peculiar "shoulder" at around 60% of the peak. This "shoulder" is evident in all the light curves, where it actually occurs as a second peak in the third burst, see fig. A1 (k, l, m). This "shoulder" is also evident in the spectral analysis of the second burst, where the bolometric flux is constant over a period of $\approx 150$ s, see fig. 13. Furthermore, we have found that this "shoulder" is evident in light curves from all the bursts, detected from this source.

# 6 DISCUSSION

## 6.1 Burst energetics

The duration of X-ray bursts is determined by the composition and thickness of the fuel layer. The total energy release, $E_b = F_p \times \tau(1 - e^{-T/\tau})$ (where we use e-folding times from table 2 as $\tau$, and T to be the time when the burst intensity stays above 20% of the peak) in order of $10^{40} - 10^{41}$ erg. Based on the observed $E_b$, we calculate an ignition depth for each of the bursts in our sample, using the formula $y_b = E_b(1+z)/4\pi R^2 Q_{nuc}$. We use $Q_{nuc} = 1.6$ MeV per nucleon, which is the energy release for burning helium to iron group nuclei. Furthermore we use $R = 10km$ (radius of the neutron star), and $z = 0.31$ (gravitational redshift), assuming a $1.4M_\odot$ neutron star. In general, the ignition column depths associated with intermediate bursts are in the order of $10^9 - 10^{10} gcm^{-2}$, assuming a thick He. The bolometric peak flux of each burst is obtained by taking the ratio of the count rate at the peak in 1s light curve of the burst and the average count rate given in the spectrum of the time-interval at the peak, and then multiplying this ratio by the bolometric flux obtained from spectroscopy of the spectrum at the peak.

**Table 2.** Properties of the bursts studied in this paper.

| Source | obs. date | Bol. $F_{peak}$ [a] | e-folding time (s) | d (kpc) | $E_b$ [b] | $y_b$ [c] |
|---|---|---|---|---|---|---|
| AX J1754.2-2754 | 2005-04-16 | $5.10 \pm 0.20$ | $56 \pm 5$ | 7.1 | $1.65 \pm 0.06$ | $1.1 \pm 0.2$ |
| AX J1754.2-2754 | 2017-03-12 | $4.9 \pm 0.25$ | $113 \pm 13$ | 7.1 | $3.2 \pm 0.1$ | $2.1 \pm 0.2$ |
| GX 3+1 | 2004-08-31 | $3.5 \pm 0.30$ | $1011 \pm 140$ | 5 | $6.3 \pm 0.3$ | $4.1 \pm 0.2$ |
| GX 17+2 | 2012-03-25 | $5.0 \pm 0.15$ | $576 \pm 120$ | 8 | $17 \pm 4.0$ | $11 \pm 3.0$ |
| GX 17+2 | 2012-08-21 | $5.7 \pm 0.20$ | $347 \pm 90$ | 8 | $16 \pm 4.0$ | $10 \pm 3.0$ |
| IGR J17254-3257 | 2006-10-01 | $1.5 \pm 0.10$ | $182 \pm 50$ | 14.5 | $9 \pm 3$ | $5.9 \pm 2$ |
| SAX J1712.6-3739 | 2018-02-20 | $8.32 \pm 0.20$ | $86 \pm 10$ | 7 | $4.9 \pm 0.3$ | $3.2 \pm 0.2$ |
| SLX 1737-282 | 2004-03-09 | $6.2 \pm 0.10$ | $230 \pm 80$ | 7.3 | $11 \pm 1.0$ | $7.2 \pm 0.7$ |
| SLX 1737-282 | 2005-04-11 | $5.7 \pm 0.25$ | $245 \pm 20$ | 7.3 | $11 \pm 1.0$ | $7.2 \pm 0.7$ |
| SLX 1737-282 | 2007-04-02 | $5.1 \pm 0.15$ | $195 \pm 15$ | 7.3 | $9.1 \pm 1$ | $5.9 \pm 0.7$ |
| SLX 1744-299 | 2013-04-06 | $4.3 \pm 0.20$ | $76 \pm 12$ | 8.5 | $2.8 \pm 0.2$ | $1.8 \pm 0.1$ |
| SLX 1744-299 | 2015-02-27 | $3.0 \pm 0.10$ | $172 \pm 30$ | 8.5 | $4.8 \pm 0.6$ | $3.1 \pm 0.4$ |
| SLX 1744-299 | 2016-03-07 | $4.0 \pm 0.15$ | $55 \pm 10$ | 8.5 | $1.90 \pm 0.3$ | $1.2 \pm 0.2$ |

[a] ($\times 10^{-8}$ erg cm$^{-2}$ s$^{-1}$)  [b] ($\times 10^{40}$ erg)  [c] $10^9$ g*cm$^{-2}$

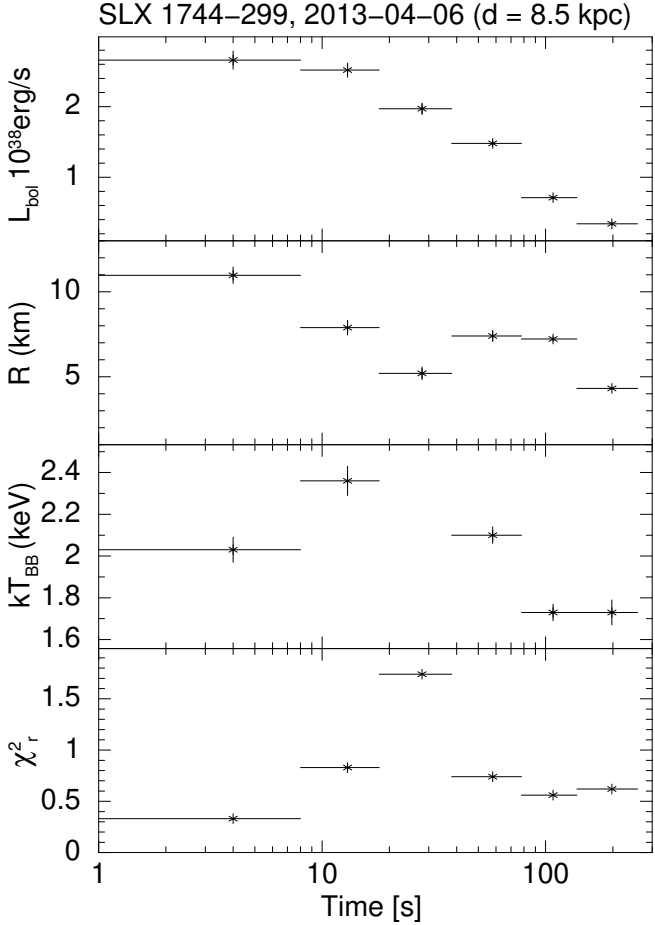

**Figure 11.** SLX 1744-299: Time-resolved spectroscopy.

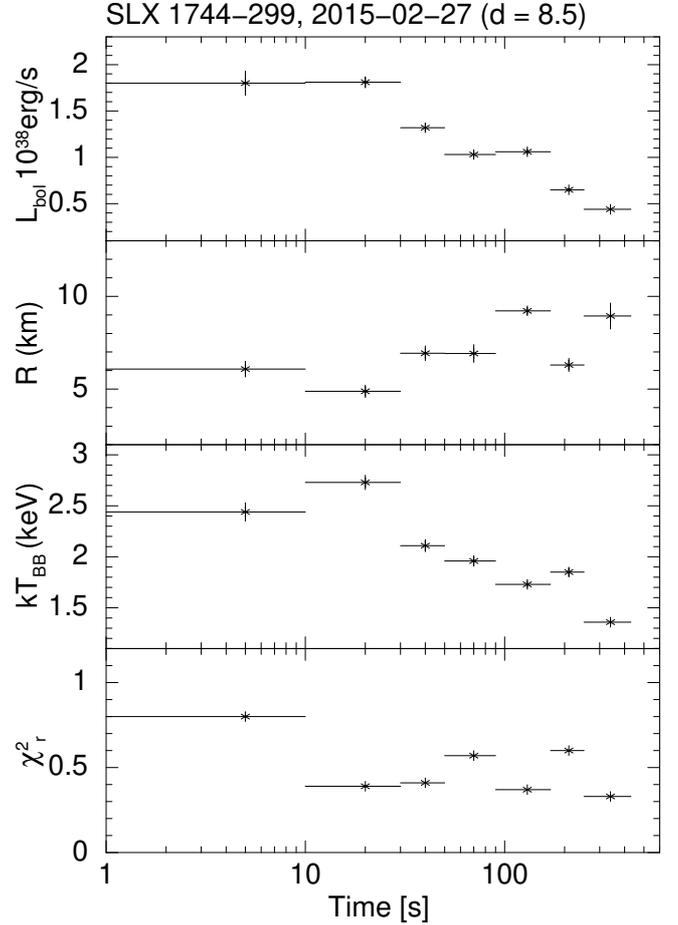

**Figure 12.** SLX 1744-299: Time-resolved spectroscopy.

We present $y_b$ and Bol. $F_{peak}$ for all the JEM-X bursts in Table 2.

All bursts detected by JEM-X exhibit the same characteristics as the intermediate bursts (i. e. $E_b$ and $y_b$). We cannot conclude anything definitive about the nature of the only burst detected by IBIS/ISGRI, since we only observe the rise phase. We note the three bursts from SLX 1744-299 having $E_b$ and $y_b$, which are on the edge of the ranges defined for intermediate bursts. Furthermore, (Galloway &

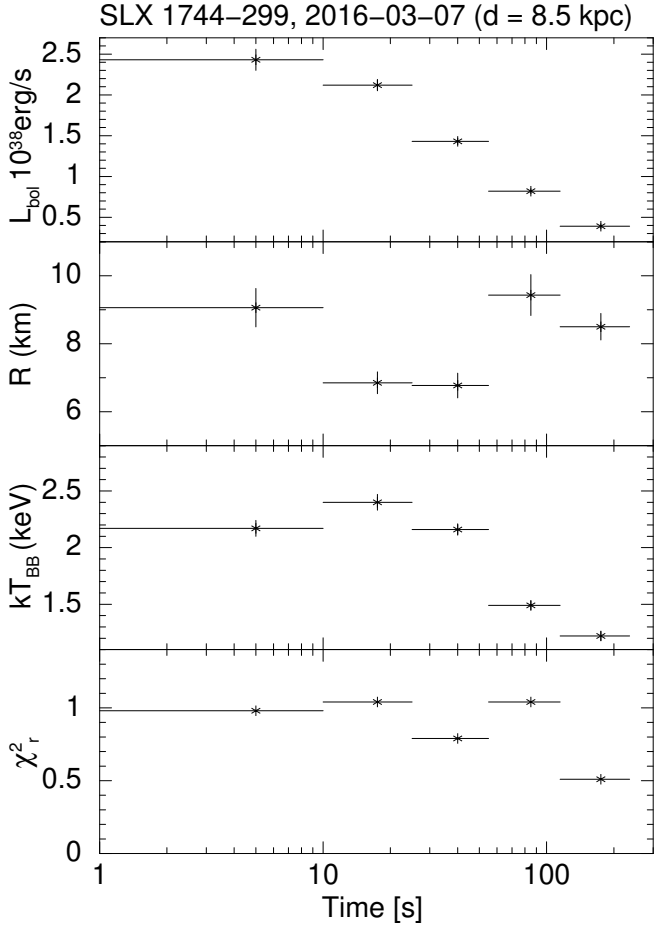

**Figure 13.** SLX 1744-299: Time-resolved spectroscopy.

Keek, 2017) state that all intermediate bursts reach the Eddington limit. This is not the case in any of the three bursts, we have observed from this source. Moreover, the bursts show a peculiar 'shoulder' at $\approx$ 60% of the peak, and have prolonged tails, which could indicate hydrogen burning through rp-process (Heger et al., 2007). The nature of SLX 1744-299 and its peculiar bursts are out of the scope of this study, but will be revisited in future work.

### 6.2 Enhanced persistent emission

The variable persistent flux method or $f_a$-method for simplicity, has been a standard method for analyzing regular Type I X-ray bursts since 2015 (Worpel et al., 2013 & Worpel et al., 2015). The procedure can be described as follows; i) a pre-burst spectrum is obtained to model the accretion mission, ii) parameters from the persistent emission model are included in the burst spectral fits and kept fixed at the values obtained from the fit of the pre-burst spectrum, iii) the persistent emission is multiplied with a variable normalization factor $f_a$. The $f_a$-method has been tested for both PRE- and non-PRE regular bursts. The physical interpretation of the $f_a$-factor is still a cause of debate, but an increase in the mass accretion rate is likely due to the effects of Poynting-Robertson drag on the disk material (Worpel et al., 2013).

Degenaar et al., 2016 applied a linear unsupervised decomposition method (non-negative matrix factorization (NMF)) to study a burst from 4U 1608-52, observed with *NuSTAR*. The main conclusion of the NMF method was a spectral softening of the persistent emission, which Degenaar et al., 2016 ascribed to the cooling of a corona.

We apply the same method to the bursts from our sample, for which we have detectable persistent emission. The TRSAs presented in section 5.8.3 are not obtained using the $f_a$-method, because the quality of our fits is not significantly improved, due to the poor spectral resolution.

Nevertheless, the $f_a$-factor shows structure with an enhanced persistent emission in the peak and tail of bursts, which is comparable to the structure of regular bursts, reported by Worpel et al., 2013; Worpel et al., 2015 & Degenaar et al., 2016. Furthermore, we see the persistent emission increase by a higher value of the $f_a$-factor, $\approx$ 22 and $\approx$ 6, for a low-accreting source like SLX 1744-299, while the high-accreting sources like GX 17+2 and GX 3+1 show a dampened increase of the persistent emission (see fig. 14). We are not able to apply the $f_a$-method on 3 of the 5 PRE-bursts in our sample, since their persistent emission level is below the threshold of the instrument. For the remaining 2 PRE-bursts, the $f_a$-factor remains nearly constant at $\approx$ 1.

The indicative structure of the $f_a$-factor, mentioned above, justifies a deeper study of the enhanced persistent emission during long-duration X-ray bursts observed in instruments at higher spectral resolution.

## 7 CONCLUSION

We have performed a systematic analysis (using the same absorption model TBABS and the same energy binning) of all Intermediate-duration X-ray bursts observed with the JEM-X instrument onboard the *INTEGRAL* satellite. With this study, we provide 13 bolometric light curves, which are suitable for comparison with numerical models. In particular, the bolometric light curves can be fitted to cooling models, and thereby providing a constraint on the energy release and ignition depth, independent of the recurrence time. This work is currently underway, and will soon be published elsewhere.

As for the eight previously reported bursts, our analysis reveal more clear indications of all three bursts from SAX 1737-282 being PRE bursts, where previously only one was identified as a PRE (Falanga et al., 2008), the first ever TRSA of the long burst from IGR J17254-3257 has been possible, and we have in general increased the number of time intervals, thereby increasing the resolution of the TRSAs.

From the new bursts, we identify one PRE with a relative strong expansion from AX J1754.2-2754, and we report three bursts from the UCXB candidate SLX 1744-299 (in 't Zand et al., 2007), which show a peculiar 'shoulder' around $\approx$ 60% of the peak and none of these show a PRE.

We apply the $f_a$ method on five of the bursts in our sample, for which we had a reasonable persistent emission spectrum. This does not improve our fits, but shows an indicative structure of the $f_a$-factor.

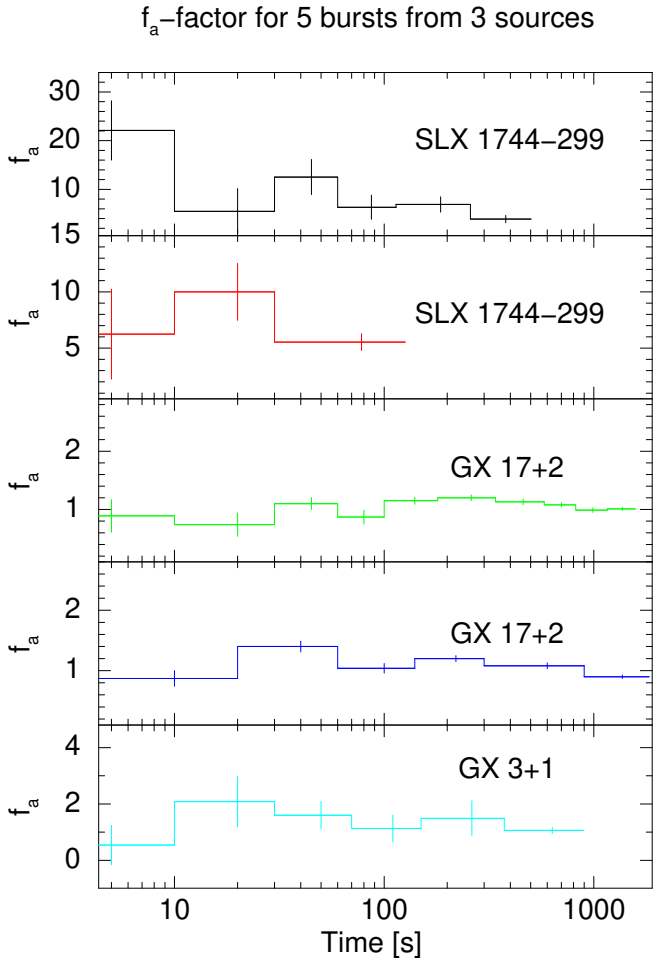

**Figure 14.** Evolution of the $f_a$-factor.

### ACKNOWLEDGEMENTS

We are grateful to all the authors of the published work, which has been reanalyzed for this study, for their support. Lastly, we would like to convey our sincere thank you to Nathalie Degenaar from the Anton Pannekoek Institute for Astronomy, University of Amsterdam, Amsterdam, Jean in't Zand from SRON, Utrecht and Erik Kuulkers from ESA/ESTEC, AZ Noordwijk. The fruitful discussions we had with both of them in the last phases of this study, have been pivotal for this work.

This work is based on the long-term observations performed by the INTEGRAL international astrophysical gamma-ray observatory and retrieved via the Russian and European INTEGRAL Science Data Centers.

## REFERENCES

Arnaud K. A., 1996, in Jacoby G.H., & Barnes J., eds, Astronomical Data Analysis Software and Systems V. ASP Conf. Series 101, p. 17
Band, D., Ferrara, E. 2009, *CICERONE: Detailed Manual for the Fermi Science Tools*, NASA/GSFC
Chelovekov, I. V., Grebenev, S. A. 2007, AL, 33, 12
Chenevez, J., Falanga, M., Brandt, S. et al. 2006, A&A, 449, L5
Chenevez J., Falanga M., Brandt S., et al. 2007, A&A, 469, L27
Chenevez J., Brandt S., Kuulkers E., et al. 2011, ATel 3183
Chernyakova, M., et al., 2015, IBIS Analysis User Manual, ISDC
Cornelisse R., Heise J., Kuulkers E., et al. 2000, A&A, 357, L21
Cumming, A., & Bildsten, L. 2000, ApJ, 544, 453
Cumming A., & Bildsten L. 2001, ApJ, 559, L127
Cumming A., & Macbeth J. 2004, ApJ, 603, L37
Cumming, A., Macbeth, J., in 't Zand, J., Page, D. 2006, ApJ, 646, 429
Degenaar, N., Koljonen, K. I. I., Chakrabarty, D., et al., 2016, MNRAS, 456, 4256-4265
Falanga M., Chenevez J., Cumming A., et al., 2008, A&A, 484, 43
Galloway D.K., Keek L. 2017, ArXiv e-prints, arXiv:1712.06227
Galloway D.K., in 't Zand J.J.M., Chenevez J. 2020, *Multi-INstrument Burst ARchive (MINBAR)* - in preparation.
Grindlay, J., Gursky, H., Schnopper, H., et al. 1976, ApJ, 205, L127
Hansen, C. J. & van Horn, H. M. 1975, ApJ, 195, 735
Heger A., Cumming A., Galloway D., and Woosley S.E. 2007, ApJ 671, L141
in 't Zand J.J.M. 1992, Ph.D Thesis, University of Utrecht
in 't Zand J.J.M., Heise, J., Bazzano A., Cocchi M., & Smith, M. J. S. 1999, IAU Circ. 7243
in 't Zand J.J.M., Cornelisse, R., Cumming A. 2004, A&A 426, 257
in 't Zand J.J.M., Cumming A., et al. 2005, A&A 441, 675
in 't Zand, J.J.M., Jonker, P. G., & Markwardt, C. B. 2007, A&A, 465, 953,
Jager, R., Mels, W. A., et al., 1997, A&A, 125, 557-572
Keek L., in 't Zand J.J.M., Kuulkers E., et al. 2008, A&A, 479, 177
Kuulkers E., van der Klis, M., Oosterbroek, T., et al. 1997, MNRAS, 287, 495
Kuulkers E., Homan, J., van der Klis, M., et al. 2002, A&A, 382, 947
Kuulkers E., 2002, A&A, 383, L5
Kuulkers E., et al. 2003, A&A, 399, 663
Kuulkers E. 2004, Nucl. Phys. B 132, 466
Kuulkers E., in 't Zand J.J.M., Atteia, J.-L., et al. 2010, A&A, 514, A65
Kuulkers E. 2011, *JEM-X Observer's Manual*, ESA
Lewin W.H.G., van Paradijs J. & Taam R. 1993, Space Science Reviews, 62, 223
Lund N., et al. 2003, A&A, 411, L231
Maraschi, L. Cavaliere, A. 1977, in X-ray Binaries and Compact objects, 999
Markwardt C.B., Altamirano D., Swank J.H., in 't Zand

J., 2008, ATel 1460

Markwardt C.B., Barthelmy, C. D., Cummings, J. C., et al., 2007, *The SWIFT BAT Software Guide* NASA/GSFC

Molkov, S., Revnivtsev, M., Lutovinov, A. & Sunyaev, R. A. 2005, A&A, 434, 1069

Oosterbroek, T., Barret, D., Guainazzi, M., et al. 2001, A&A, 366, 138

Pavlinsky, M. N., Grebenev, & Sunyaev, R. A. 1994, ApJ, 425, 110-121

Sakano, M., Koyama, K., et al., 2002, ApJ Suppl. Ser. 138, 19

Savchenko, V., Chernyakov, M., et al., 2018, *The IBIS BAT Analysis User Manual* ISDC

Seifina, E., Titarchuk, L. 2012, ApJ, 747, 99

Sguera, V., Bazzano, A., Bird, A. J. 2007, ATel 1340

Skinner, G. K., Willmore, A. P., Eyles, C. J., et al. 1987, Nature, 330, 544

Strohmayer T.E.,& Bildsten L. 2006, in Compact stellar X-ray sources,

Swank, J. H., Becker, R. H., Boldt, E. A., et al. 1977, ApJ, 212, L73

Walter, R., Bodaghee, A., Barlow, E. J. 2004, ATel 229

Werner N., in 't Zand J., Natalucci L., et al. 2004, A&A 416, 311

Westergaard, N. J., Kretschmar, P., Oxborrow, C. A., et al., 2003, A&A 411, L257-L260

Wijnands R. 2001, ApJL, 554, L59

Worpel, H., Galloway, D., Price, D. J. 2013, ApJ 772, 94

Worpel, H., Galloway, D., Price, D. J. 2015, ApJ, 801, 60

# APPENDIX A: LIGHT CURVES

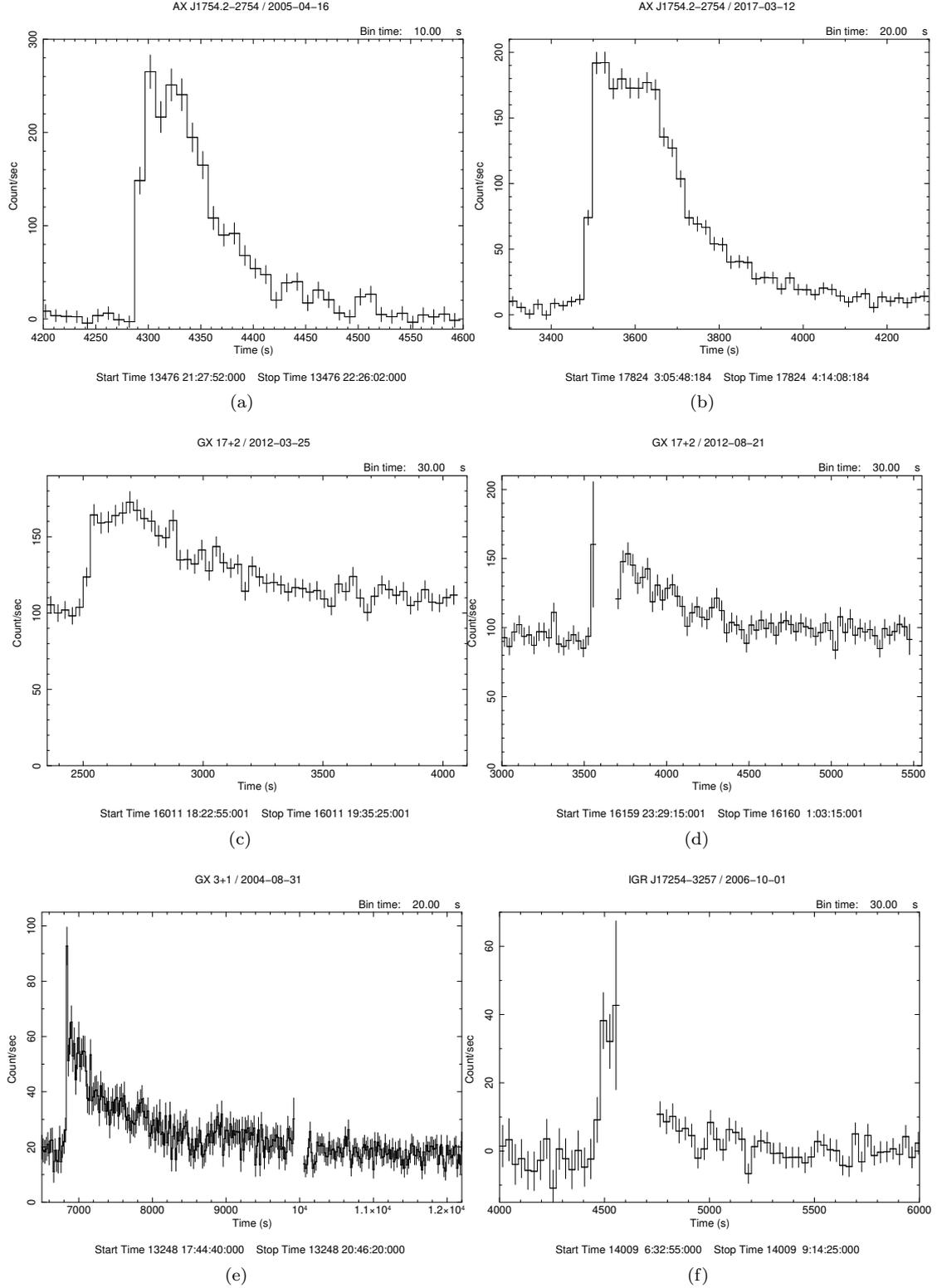

**Figure A1.** Lightcurves of the X-ray bursts investigated in this study.

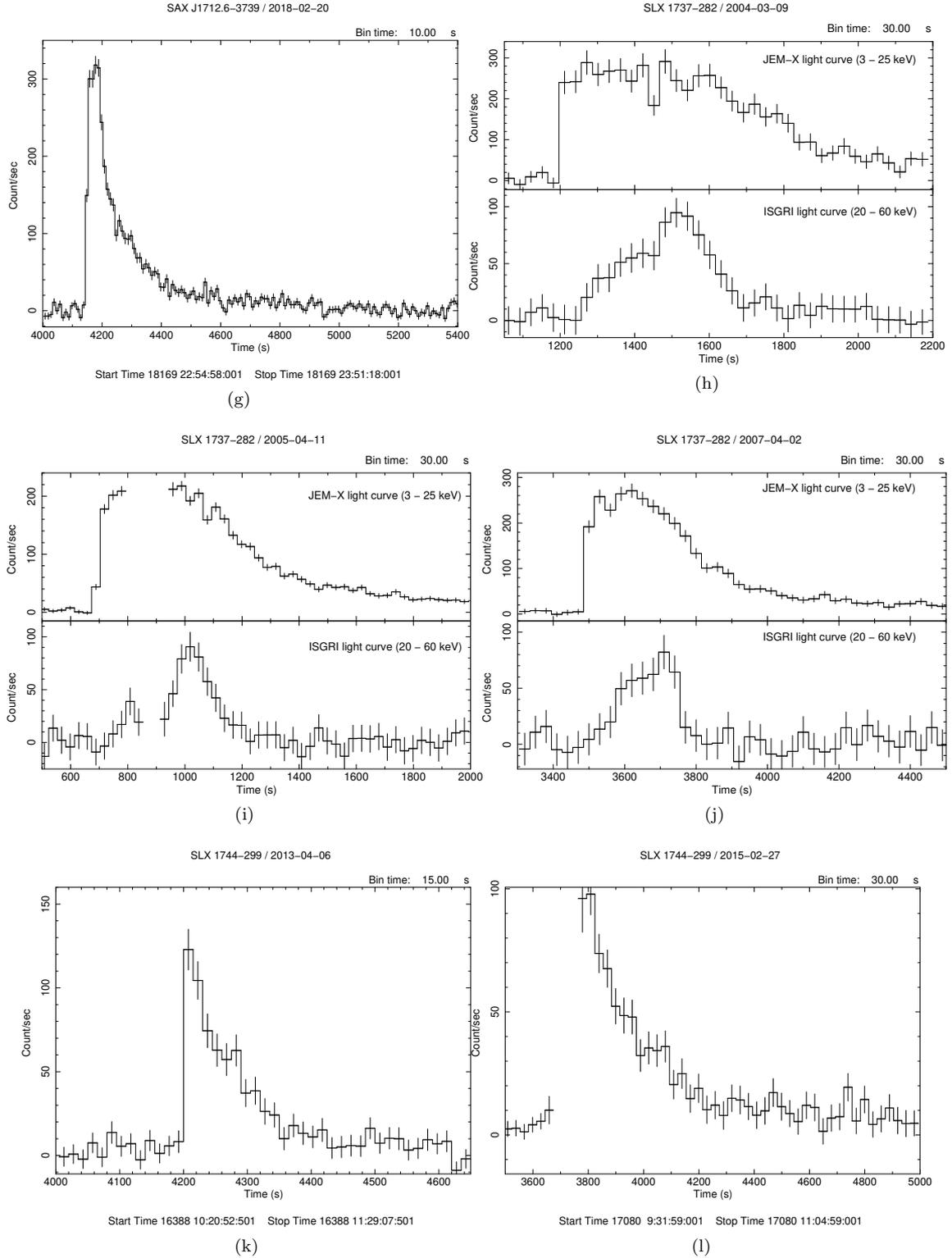

**Figure A1.** Lightcurves of the X-ray bursts investigated in this study.

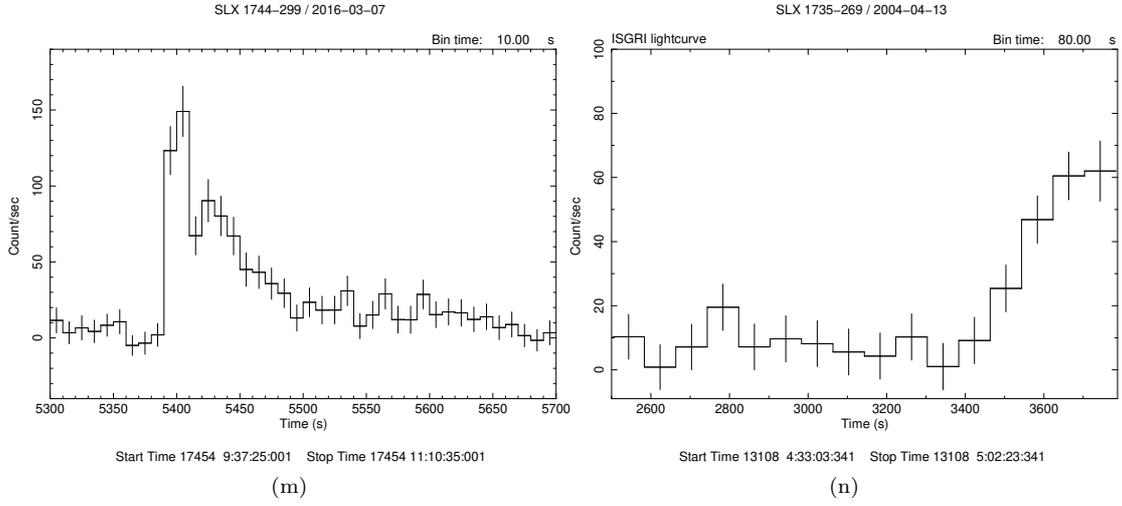

**Figure A1.** Lightcurves of the X-ray bursts investigated in this study.